\PassOptionsToPackage{table,prologue,dvipsnames}{xcolor}
\documentclass[]{bytedance_seed}

\usepackage[toc,page,header]{appendix}

\usepackage{minitoc}

\usepackage{amsfonts}
\usepackage{amsmath}
\usepackage{amssymb}
\usepackage{lineno}
\usepackage{wrapfig}  
\usepackage{booktabs} 
\usepackage[bottom]{footmisc}

\usepackage{CJKutf8}
\usepackage{setspace}
\usepackage{arydshln}
\usepackage{xspace} 

\usepackage{graphicx}
\usepackage{subcaption}
\usepackage{xcolor}
\usepackage{colortbl}

\usepackage{algorithm}
\usepackage{algorithmicx}
\usepackage{algpseudocode}

\definecolor{linkcolor}{rgb}{0.956,0.298,0.235}
\definecolor{citecolor}{HTML}{1976D2}
\hypersetup{
    colorlinks=true,    
    linkcolor=citecolor,     
    filecolor=magenta,  
    urlcolor=cyan,      
    citecolor=citecolor,    
}

\renewcommand{\paragraph}[1]{\noindent\textbf{#1.}\hspace*{1em}}
\definecolor{customblue}{HTML}{DAE8FC}
\definecolor{customyellow}{HTML}{FFF2CC}

\renewcommand{\algorithmicrequire}{\textbf{Input:}}
\renewcommand{\algorithmicensure}{\textbf{Output:}}

\title{DreamFoley: Scalable VLMs for High-Fidelity Video-to-Audio Generation}
\author[~1,*]{Fu Li}
\author[~1,*]{Weichao Zhao}
\author[~1,2,*]{You Li}
\author[~1]{Zhichao Zhou}
\author[~1,\dagger]{Dongliang He}

\affiliation[1]{Bytedance Intelligent Creation Lab}
\affiliation[2]{Zhejiang University}

\contribution[*]{Equal Contribution}
\contribution[\dagger]{Project Lead}

\abstract{
Recent advances in video generation have achieved remarkable improvements in visual content fidelity.
However, the absence of synchronized audio severely undermines immersive experience and restricts practical applications of these technologies.
To address this challenge, several pioneering works have explored diffusion transformer architectures for generating plausible video-synchronized audio, including \textit{Kling-foley}~\cite{wang2025kling}, \textit{HunyuanVideo-foley}~\cite{shan2025hunyuanvideo} and \textit{Thinksound}~\cite{liu2025thinksound}.
Distinct from existing works, we introduce an autoregressive audio generation architecture~(DreamFoley) that harnesses the capabilities of large vision-language models (VLMs) to jointly model sequential interactions among video, audio, and text modalities.
Our approach features a dual-visual encoder module that effectively captures both audio-aligned and text-aligned visual features.
Additionally, we employ a Residual Vector Quantization audio tokenizer with a delay-pattern generation scheme to balance the trade-off between training efficiency and audio quality.
Moreover, we introduce the classifier-free guidance strategy into VLMs to bootstrap generated audio quality.
Furthermore, we establish an efficient data production pipeline to scale audio-video-text triple collection.
Finally, extensive experiments are conducted to validate the effectiveness of our model, achieving promising performance across popular benchmarks. 
We hope that the findings in this study provide a strong foundation for future video-to-audio generation research.
We also release the previously missing audio-visual textual descriptions from the public benchmark, aiming to facilitate subsequent researchers in conducting more convenient and effective evaluations and comparisons.
}
\date{\today}

\checkdata[Project Page]{\href{https://sakura2233565548.github.io/DreamFoley}{https://Bytedance/DreamFoley}}
\checkdata[Benchmark]{\href{https://huggingface.co/datasets/Zhaowc/DreamFoley}{https://huggingface.co/datasets/DreamFoley}}

\begin{document}
\maketitle






\section{Introduction}
\label{sec:intro}

Video generation technology has emerged as one of the most prominent research directions in the field of general artificial intelligence. 
Recent years have witnessed remarkable progress in both open-source models (such as Wan2.1~\cite{wan2025wan} and HunyuanVideo~\cite{kong2024hunyuanvideo}) and closed-source models (such as the Seedream~\cite{seedream2025seedream} and Kling series~\cite{klingvideo}) in generating realistic video content. 
However, despite these advancements, generated videos typically remain silent and lack synchronized audio information, which significantly diminishes user experience and limits practical applications. 
Manual dubbing requires specialized technical expertise, substantial time and financial investment, severely constraining the efficiency and scalability of video generation systems.

To address the audio generation challenge, text-to-audio~(T2A) technology has evolved to automate the generation of non-speech audio content, including sound effects and background music, based on textual descriptions~\cite{kreuk2022audiogen,huang2023make,deepanway2023text,audioldm2,yang2023diffsound,zhao2025audioturbo}. 
The primary objective of T2A is to translate textual information into semantically aligned, high-fidelity audio sequences. 
However, textual descriptions inherently lack fine-grained temporal alignment with precise sound onset moments in video content, frequently resulting in mismatches between generated sound effects and the actual video events.

Video-to-audio (V2A) approaches have recently incorporated both textual and visual information to address these limitations~\cite{cheng2025mmaudio,shan2025hunyuanvideo,wang2025audiogen,liu2025thinksound,wang2025kling}. 
MMAudio~\cite{cheng2025mmaudio} represents a pioneering work in this domain, introducing a flow-matching architecture that leverages optional multimodal conditions to generate high-fidelity monaural audio. 
Subsequent studies have enhanced audio quality through various strategies, including dataset scaling~\cite{wang2025kling,wang2025audiogen}, textual descriptions refinement~\cite{liu2025thinksound}, and architecture improvement~\cite{shan2025hunyuanvideo}.
Despite these promising developments, current methods predominantly rely on the flow-matching paradigm, which presents several critical limitations:
1) \textbf{Temporal Scalability}. Previous flow-matching models are typically trained on fixed-length video sequences, constraining them to approximately 10-second durations. When applied to longer videos, these approaches experience significant quality degradation as they struggle to capture complex audio dynamics in extended temporal sequences.
2) \textbf{Semantic Alignment}. Previous works embed textual audio descriptions using pre-trained language models~\cite{yang2025qwen2,chung2024scaling,touvron2023llama}, and integrate semantic features through cross-attention mechanisms. 
However, this paradigm does not inherently guarantee semantic consistency between textual and audio modalities, resulting in discrepancies between intended textual meaning and generated audio content.

To address these challenges, we propose Dream-Foley, a novel autoregressive framework that leverages the comprehensive capability of Vision-Language Models (VLMs) to generate synchronized, high-quality audio for video content.
Our framework centers on a large language model that serves as a multimodal bridge, integrating information across modalities at the semantic level.
Specifically, we introduce a dual-vision encoder architecture where each encoder is independently pre-trained on textual and visual modalities, respectively, enabling extraction of critical cues for plausible audio generation and precise audio-visual synchronization. 
Moreover, to align with the next-token prediction paradigm of large language models (LLMs), we employ a discrete tokenizer that maps original audio signals into discrete embedding spaces, facilitating seamless integration with autoregressive modeling.
Since quantization in discrete audio tokenization process often causes semantic information loss, we employ Residual Vector Quantization (RVQ) combined with Multi-Token Prediction (MTP) to mitigate semantic degradation and enhance audio representation quality.
Inspired by the advances of classifier-free guidance~(CFG) strategy, we successfully integrate this scheme into our framework without introducing any additional architectural components and validate its positive impact on generated audio quality.
Finally, we establish an efficient data collection strategy to curate high-quality audio-video-text triplets for effective model training.
Extensive experiments validate our framework's effectiveness on several popular benchmarks, achieving state-of-the-art performance compared to the latest open-source methods.

\section{Related Works}

\noindent \textbf{Text-to-Audio.} 
Text-to-audio technologies have made significant progress in recent years, with applications ranging from text-to-speech (TTS) systems for natural language communication to more advanced text-to-sound and text-to-music synthesis. These technologies utilize deep learning techniques, particularly sequence-to-sequence and generative models, to synthesize audio that corresponds to textual input. 
Early models, such as AudioGen~\cite{kreuk2022audiogen}, leveraged autoregressive transformers with discrete audio representations for audio generation. CosyVoice~\cite{cosyvoice} introduced a multilingual speech synthesis framework based on supervised semantic tokens, with successive versions, CosyVoice 2~\cite{cosyvoice2} and CosyVoice 3~\cite{cosyvoice3}  improving real-time streaming, prosody, and speaker similarity through scalable data and innovative tokenization methods. MAGNET~\cite{magnet} proposed a masked generative transformer that directly models audio tokens with efficient parallel decoding, significantly improving both generation speed and quality.
Meanwhile, the emergence of diffusion models has introduced a powerful paradigm for audio synthesis. Early works like Diffsound~\cite{yang2023diffsound} demonstrated the effectiveness of discrete diffusion decoders in generating high-quality sounds from text. This approach evolved with latent diffusion models (LDMs)~\cite{ldm}, which operate in compressed audio representations to improve computational efficiency. AudioLDM~\cite{audioldm} incorporated CLAP embeddings~\cite{wu2023large} to guide the generative process, while its successor, AudioLDM 2~\cite{audioldm2}, further enhanced the framework with self-supervised AudioMAE~\cite{audiomae} features, creating a more unified and capable system. TANGO~\cite{tango} leveraged an instruction-tuned FLAN-T5~\cite{chung2024scaling} encoder to strengthen language understanding, and Make-An-Audio~\cite{makeanaudio} combined CLAP guidance with spectrogram autoencoding to achieve more controllable and high-fidelity generation. The introduction of TANGOFLUX~\cite{tangoflux} integrated a hybrid Diffusion Transformer (DiT) architecture with CLAP-ranked preference optimization, further refining output quality, aligning generation with human preferences, and significantly reducing inference latency.

\noindent \textbf{Video-to-Audio.} 
Video-to-audio research focuses on synthesizing or enhancing audio based on visual cues and contextual information extracted from video input. 
FoleyCrafter~\cite{zhang2024foleycrafter} leveraged a pre-trained AudioGen transformer to generate audio conditioned on visual inputs. Diff-Foley~\cite{difffoley} introduced a latent diffusion-based V2A framework that takes advantage of contrastive audio-visual pretraining into the audio generation process. Frieren~\cite{wang2024frieren} further improved the generation efficiency through rectified flow matching and cross-modal fusion, achieving high-quality audio with near one-step sampling. To better understand multimodal representation learning, VTA-LDM~\cite{vta-ldm} systematically explored the roles of visual encoders, auxiliary embeddings and data augmentation strategies, while V-AURA~\cite{viertola2025temporally} employed high-framerate visual feature extraction and cross-modal fusion within an autoregressive framework for superior synchronization. Expanding the controllability of generation, MultiFoley~\cite{multifoley} enabled flexible sound design conditioned on video, text, and audio input. MMAudio~\cite{cheng2025mmaudio} trained together on large-scale text-audio and video data to improve both semantic alignment and inference efficiency. In addition, ThinkSound~\cite{liu2025thinksound} introduced Chain-of-Thought (CoT) reasoning for stepwise, interactive, and cognitively guided audio synthesis. More recently, large-scale multimodal frameworks such as HunyuanVideo-Foley~\cite{shan2025hunyuanvideo} and Kling-Foley~\cite{wang2025kling} employed diffusion transformers with multimodal fusion and synchronization modules, achieving state-of-the-art performance in audio fidelity, semantic alignment, and fine-grained temporal correspondence.
\section{Model Architecture}
\label{sec:model-arch}

\subsection{Overview}
\begin{figure}[t]
	\centering
	\includegraphics[width=0.95\linewidth]{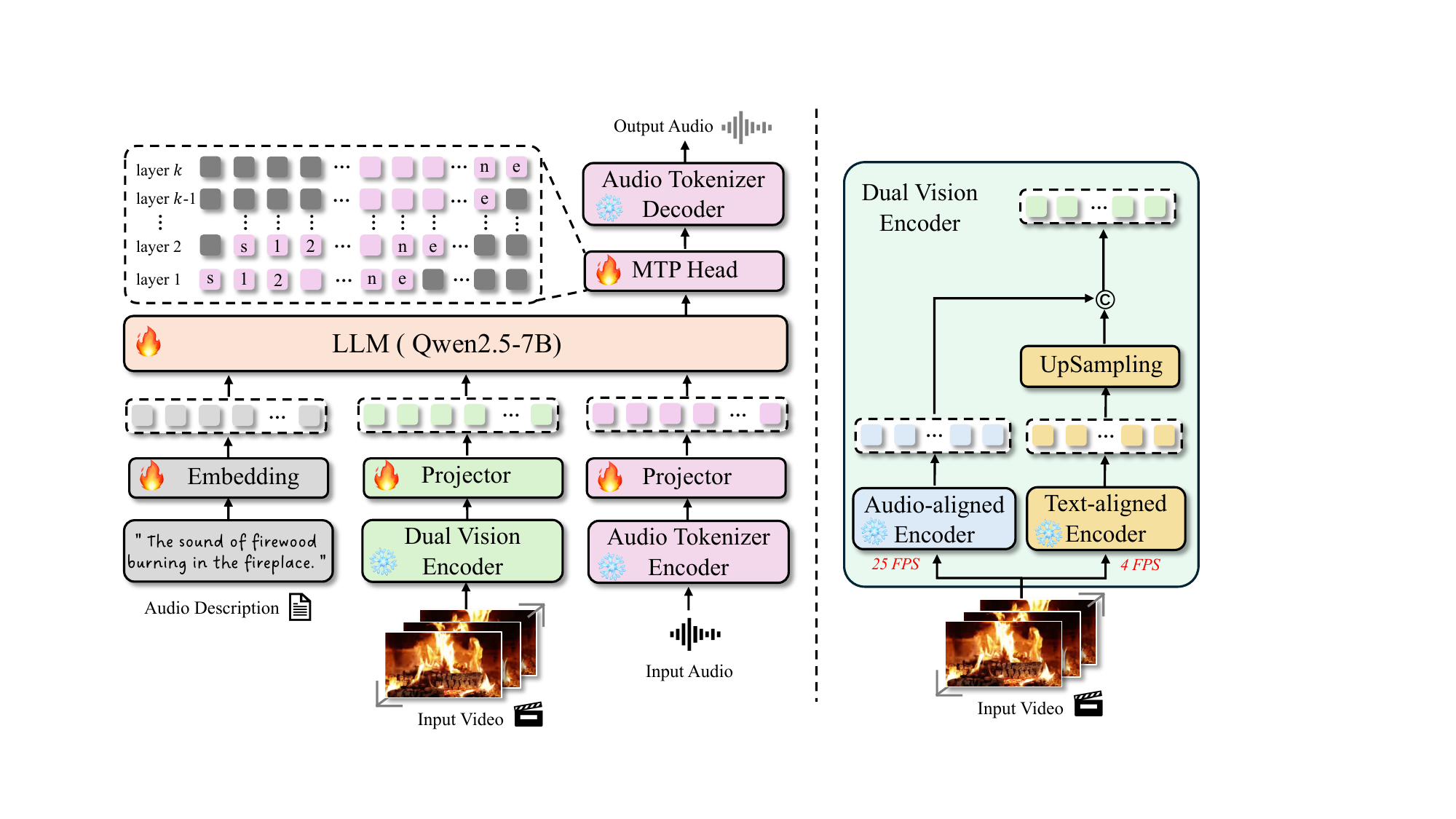} 
	\caption{Overview of our proposed framework. DreamFoley mainly includes three components, \textit{i.e.,} the dual vision encoder, the audio tokenizer and the LLM. The dual vision encoder consists of audio-aligned and text-aligned encoders.} 
	\label{fig:framework}
	\vspace{-1.2em}
\end{figure}

In this section, we provide the details of our proposed framework.
As illustrated in Fig.~\ref{fig:framework}, our framework is built on the decoder-only LLM model, i.e., Qwen2.5-7B~\cite{yang2025qwen2p5}.
By leveraging its advanced capabilities to capture long-range contextual dependencies, the model inherently learns the flow of causal information during the training process. 
To accommodate diverse data types, we employ modality-specific encoders that extract and embed information from multiple modalities. This information is then integrated into the large language model (LLM). This architectural design allows the model to effectively process and synthesize sequential information across modalities, facilitating the generation of high-fidelity audio signals.
Next, we will elaborate on several crucial components of our model.

\subsection{Model Components}
\noindent \textbf{Dual Vision Encoder.} 
Given an input video $V_{in}$ with duration $T$, we employ a dual vision encoder to extract complementary visual representations, as shown on the right side of Fig.~\ref{fig:framework}.
This architecture consists of two specialized branches designed to capture distinct but complementary visual information.

The first branch, termed the audio-aligned encoder~$\psi_a$, is pre-trained on paired audio-video clips using a CLIP-style contrastive learning strategy. 
Since audio events often exhibit rapid temporal variations, this encoder operates at a high video sampling rate $r_a$ to capture fine-grained temporal alignment between audio and visual events. The increased sampling frequency provides higher temporal resolution, enabling our model to effectively synchronize fast-changing audio cues with the corresponding visual dynamics.

The second branch, referred to as the text-aligned encoder~$\psi_t$, is trained on paired text-image data within a contrastive learning framework. 
This encoder focuses on capturing global visual semantics, providing complementary information to the temporally focused features extracted by $\psi_a$.
Since visual scene content in videos typically evolves more gradually than audio events, the text-aligned encoder operates at a lower sampling rate $r_t$, effectively balancing visual comprehension with computational efficiency. 

Finally, we obtain the visual features of both branches, denoted as $F_a$ and $F_t$ respectively.
Due to the different sampling rate, we apply an upsampling operation $\Psi_{up}$ to the text-aligned encoder output to ensure consistent feature length. 
The final visual representation is formulated as:
\begin{equation}
	F_v = concat(\psi_t(V_{in}, r_a), \Psi_{up}(\psi_a(V_{in}, r_t), \lceil r_a/r_t \rceil)).
\end{equation}
In our framework, we adopt EVA-CLIP-g/14~\cite{sun2023eva} and Synchformer~\cite{iashin2024synchformer} as the text-aligned and audio-aligned encoders, respectively.

\textbf{Audio Tokenizer.} For input audio $A_{in}$, acoustic audio codec is often used to directly embed continuous audio signals into discrete acoustic tokens.
However, standard acoustic codecs are designed primarily for audio compression without considering semantic information requirements for LLM.
To address this limitation, we adopt a codec that incorporates semantic features from a pre-trained semantic encoder.
To further enhance the representation of subtle tonal variations and preserve high-fidelity audio quality, we employ the RVQ audio codec.
This codec reconstructs audio content in a progressive manner, transitioning from coarse-grained to fine-grained detail levels.

Specifically, we utilize the higgs-audio-v2-tokenizer~\footnote{\href{https://github.com/boson-ai/higgs-audio/blob/main/tech_blogs/TOKENIZER_BLOG.md}{https://github.com/boson-ai/higgs-audio/blob/main/tech\_blogs/TOKENIZER\_BLOG.md}}, which utilizes $K=8$ codebooks to perform residual quantization of latent audio features, converting them into discrete audio tokens.
This tokenizer is jointly trained on speech, music, and sound-event data, capturing both sementic and acoustic nuances, and runs at just 25 tokens per second, therefore greatly simplifying downstream audio-language-model training.
For output audio generation, the corresponding tokenizer decoder reconstructs the audio signal $\hat{A}_{out}$ from the generated audio tokens, ensuring both fidelity and semantic consistency in the reconstructed output.
The complete audio encoding process is summarized in Algo.~\ref{algo:audio encoding}.

\textbf{Projectors.} For the text, video, and audio modalities, we employ modality-specific projection modules to ensure compatibility with the input dimensions of the LLM.
Specifically, the text modality is first tokenized using Byte Pair Encoding (BPE), and subsequently encoded into textual features through word embedding.
The video features~$F_v$ are projected with a single MLP network.
For audio tokens $R_a$, we independently initialize $K$ embedding matrices to encode audio tokens from different layers. These features are then aggregated from all layers with a delay-pattern scheme to form the final audio representation. The projecting process is formulated as follows.
\begin{align}
    H_{s} = \mathrm{WordEmbed}(S_{in}), \quad H_v = \mathrm{MLP}(F_v), \quad H_a = \sum\nolimits_{i=1}^{K}\mathrm{AudioEmbed}_{i}(\mathrm{Delay}(R_i)), R_i \in R_a
\end{align}

\begin{figure}[t]
	\centering
	\includegraphics[width=0.90\linewidth]{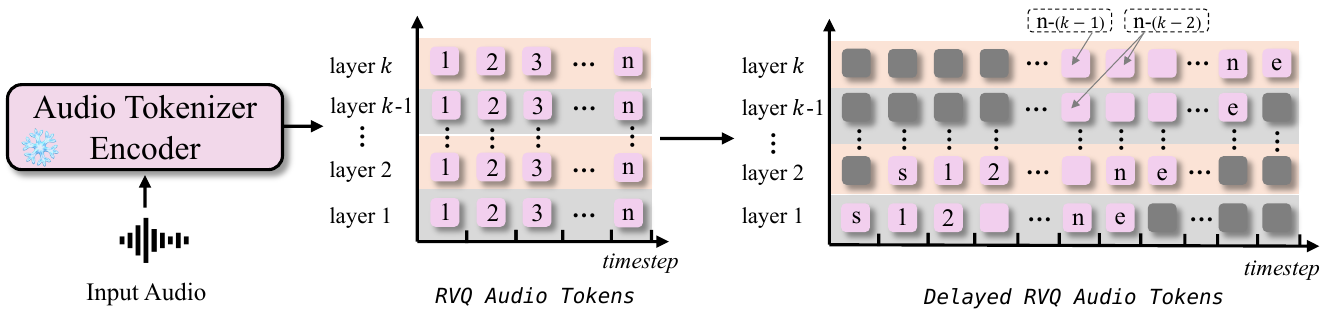} 
	\caption{Illustration of the delay generation scheme.} 
	\label{fig:delay_pattern}
\end{figure}

\vspace{-1.0em}
\begin{algorithm}[t]
    \renewcommand{\algorithmicrequire}{\textbf{Input:}}
    \renewcommand{\algorithmicensure}{\textbf{Output:}}
    \caption{Audio Encoding Algorithm}
    \label{algo:audio encoding}
    \begin{algorithmic}
        \Require $A_{in} \in \mathrm{R}^{T \times 1}$, RVQ Codebooks \{$C_1, ...,C_K$\}
        \State Initial: ${F}_a = Enc\_audio(A_{in})$, $r_0 = {F}_a$
        \Comment{Encode raw audio into latent features ${F}_a \in  \mathrm{R}^{L \times d}$}
        \For {$i = 1$ \textbf{to} $L$}
        \For{$j = 1$ \textbf{to} $K$}
            \State $k_{j,i} \gets \arg\min_{k} || r_{j-1, i} - c_{j, k} ||$ \Comment{Find the nearest index in $C_j = \{c_{j, k}\}_{k=1}^{D}$}
            \State $r_{j, i} \gets r_{j-1, i} - c_{j, k_{j,i}}$ \Comment{Update residual feature}
        \EndFor
        \EndFor
        \State Aggregate: $R_j = \{k_{j,1}, \cdots, k_{j, L}\}$ \Comment{Aggregate the index sequence of the $j$-th codebook}
    \Ensure $R_a = \{R_1, \cdots, R_K\} $
    \end{algorithmic}
\end{algorithm}
\vspace{-0.5em}

\textbf{Multi-token Prediction.} 
To address the requirement of decoding multiple tokens in parallel during audio generation, we employ a delayed generation mechanism~\cite{copet2023simple} to ensure that the audio tokens are generated in a causally coherent sequence, as shown in Fig.~\ref{fig:framework}.
Concretely, causal generation encompasses two principal components. Firstly, for single-layer audio token sequences, the model predicts tokens sequentially according to their temporal order. Secondly, for multi-layer residual audio token sequences at each generation step, we utilize the residual quantization mechanism employed in the tokenizer. In this process, the audio tokens in each subsequent layer depend on the predictions from earlier layers, facilitating effective information transmission and optimization across layers.
Additionally, for the vacant positions resulting from the shifting operations across different layers, we employ learnable padding tokens to fill these gaps, ensuring sequence consistency and enhancing model training.
For clarity, we illustrate the delay pattern pipeline in Fig.~\ref{fig:delay_pattern}.

\textbf{Training Objective.} 
Since the MTP scheme requires simultaneous prediction of audio token distributions across multiple layers, we design $K$ independent MLP heads to map the output features of the LLM to each respective layer.
During training, vanilla cross-entropy loss is employed to supervise the prediction of audio tokens for each layer.
In practice, we discover that audio features from shallow layers have a greater impact on the quality of the generated audio.
In response to this finding, we implement a progressively decaying weighting scheme that differentially weights the loss values across layers, thereby improving the overall performance.

\textbf{Classifier-free Guidance~(CFG).} CFG~\cite{ho2022classifier} is a pivotal technique in diffusion-based models, known for enhancing both the quality and controllability of generated samples. 
It achieves this by reinforcing the consistency between the generated outputs and conditional inputs, such as text prompts or visual contents, during the denoising process.
Inspired by this, we extend the core principles of CFG to our AR-based framework.
Specifically, 
at each step of prediction, the target token is sampled by combining conditional and unconditional probability distributions.
The unconditional distribution is modeled using sequences of learnable embeddings for video and text inputs.
To achieve this, we adopt a random masking strategy for video and text inputs, and introduce corresponding learnable embeddings during training, thereby guiding the model to generate audio under incomplete information.
The experimental results in Sec.~\ref{sec:exp} demonstrate that our proposed strategy excels in the audio generation task, effectively improving the fidelity of the target timbre while significantly suppressing background noise. 
These findings validate the efficacy of this cross-domain application.

\section{Data Construction}
\label{sec:data}

Recent advances in large-scale model training have demonstrated that achieving strong performance requires massive and high-quality training data.
However, the V2A task suffers from performance degradation due to both data scarcity and suboptimal data quality.
Current open-source datasets~\cite{chen2020vggsound,gemmeke2017audio} typically contain only hundreds of thousands of training videos, which is insufficient for fully unleashing the potential of large-scale models.
Moreover, these datasets frequently exhibit misalignment between visual content and corresponding audio annotations, alongside audio samples of inadequate quality. These limitations introduce noise and ambiguity into the training process, negatively impacting the effectiveness of model learning.

\begin{figure}[t]
	\centering
	\includegraphics[width=1.0\linewidth]{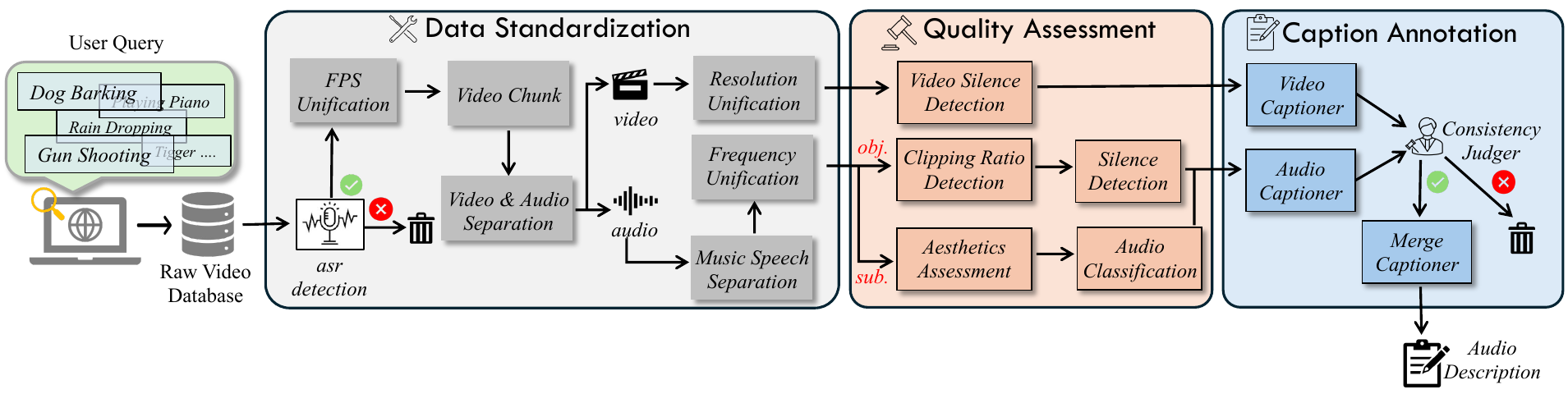} 
	\caption{Illustration of data production pipeline. The whole pipeline consists of three main components: data standardization, {quality assessment} and {caption annotation}.} 
	\label{fig:data_pipeline}
\end{figure}

To tackle these challenges, we construct an efficient data production pipeline as shown in Fig.~\ref{fig:data_pipeline}.
The whole pipeline consists of three main components: \textit{data standardization}, \textit{quality assessment} and \textit{caption annotation}.
Next, we will present each component in detail.

\subsection{Data Standardization}
This stage standardizes heterogeneous video data to establish a consistent and unified data format for subsequent processing.
We construct numerous audio related queries to retrieve diverse video data from various sources.
For each retrieved video, Automatic Speech Recognition (ASR) techniques are first applied to detect the presence of substantial human speech.
Since the V2A task primary focuses on generating pseudo sound effects rather than speech, videos containing predominantly human speech are filtered out to improve the utility of the final dataset. 
Subsequently, videos of varying durations are chunked into 10-second clips, and the audio and video streams are separated for independent processing.

Video clips are resized to 360px on the short side while preserving the original aspect ratio.
For the audio processing, since ASR detection alone is insufficient to fully exclude videos containing human speech, the music-speech separation~(MSS) technique is applied to remove residual speech signals from the audio track.
Afterwards, the audio sampling rate is standardized to 44.1kHZ.
This procedure yields a substantial collection of standardized audio-visual clips, which serve as the foundation for subsequent quality assessment.

\subsection{Quality Assessment}
This stage aims to measure the quality of video and audio from both subjective and objective perspectives.
For video clips, we examine whether the clips contain an excessive number of static shots, which may indicate lower visual engagement.
Regarding audio assessment, objective metrics include the detection of clipping ratio and periods of silence. Specifically, any sample exhibiting more than 80\% silence or exceeding a 10\% clipping ratio is excluded to maintain high-quality standards.
Subjective evaluation involves both aesthetic quality ratings and audio content classification. In particular, we employ PAM~\cite{deshmukh2024pam}, which leverages CLAP~\cite{wu2023large} as its foundational model to assess the overall perceptual quality of the samples. 
Furthermore, AST~\cite{gong2021ast} is utilized to categorize audio content into 527 distinct classes to balance data distribution. 
Samples that receive a PAM score below 0.3 are also discarded to further ensure the integrity of the dataset.
Additionally, we employ ImageBind~\cite{girdhar2023imagebind} as an evaluation tool to rigorously assess audio-video alignment.
Through the above multi-dimensional screening, we retain high-quality audio-video pairs for final caption annotation.

\subsection{Caption Annotation}
In this stage, we first caption audio and video with Qwen2.5-Omni~\cite{xu2025qwen2} and Doubao1.5-VL~\cite{guo2025seed1}, respectively.
However, we found that captions generated solely from audio often lack accuracy due to the absence of complementary visual information, which in turn affects the overall caption quality.
As a result, we design an audio-video caption consistency discriminator that verifies whether the main content described by the audio caption aligns with that of the video caption. Inconsistent samples are discarded, while consistent samples proceed to caption integration, where information from both modalities is combined to generate the final audio description with Doubao1.6~\footnote{\href{https://seed.bytedance.com/en/seed1_6}{https://seed.bytedance.com/en/seed1.6}}.

Leveraging this efficient data pipeline, we expand the training audio-video-text pairs to the scale of millions, thereby supporting generalized model training.

\section{Experiments}
\label{sec:exp}

\subsection{Implementation Details}
For the dual-vision encoder, the sampling rates $r_a$ and $r_t$ are set to 25 and 4, respectively. In accordance with MMAudio~\cite{cheng2025mmaudio}, the sliding window size and stride step in the audio-aligned encoder are set to 16 and 8, respectively.
During training, DreamFoley is conducted with 
a total batch size of 1024 over $40k$ steps, using the AdamW optimizer with a learning rate of $1e-4$.
We randomly crop small video clips ranging from 2 to 10 seconds in length. 
The generation of longer audio will be addressed in future work.
For CFG, the real video embeddings and text embeddings are replaced by empty learnable embeddings 10\% of the time respectively.
The loss weights of the training objective on different layers are set to $[0.240, 0.192, 0.154, 0.123, 0.099, 0.079, 0.063, 0.050]$.
For inference, the $temperature$, $top\_p$ and $top\_k$ in LLM are set to $1.0$, $0.9$ and $25$, respectively.

\subsection{Evaluation Metrics}
To achieve a comprehensive assessment of the quality of the generated audio, we performed an in-depth and systematic analysis across five distinct dimensions.
1) \textbf{Distribution matching} is employed to evaluate the degree of similarity between the feature distributions of generated audio and ground-truth audio within the specific embedding model. 
Concretely, we compute Fr$\mathrm{\acute{e}}$chet Distance~(FD) and Kullback–Leibler~(KL) distance.
For FD, we adopt PassT~\cite{koutini2022efficient} and PANNs~\cite{kong2020panns}
as embedding models, denoted as $\mathrm{FD_{PassT}}$, $\mathrm{FD_{PANNs}}$ and $\mathrm{FD_{VGG}}$, respectively.
For KL, we also adopt PANNs and PassT as classifiers, and report the average of $\mathrm{KL_{PANNs}}$ and $\mathrm{KL_{PassT}}$.
2) \textbf{Audio quality} assesses the quality of audio generation without the reference ground-truth.
We utilize the inception score~\cite{salimans2016improved}~(IS) to evaluate the diversity and quality of the generated audios.
3) \textbf{Semantic alignment} assesses the similarity between the generated audio and the input video.
We adopt ImageBind~\cite{girdhar2023imagebind} to extract video and audio features for each clip and compute the average cosine similarity as ``IB-Score''.
4) \textbf{Temporal alignment} assesses the audio-visual synchrony via computing a synchronization score~(DeSync) with Synchformer~\cite{iashin2024synchformer}.
5) \textbf{Text-audio consistency} evaluates the similarity between the audio description and the corresponding generated audio using LAION-CLAP~\cite{wu2023large}.


\subsection{Comparison with State-of-the-art Methods}
In this section, we compare the DreamFoley with previous methods on VGGSound-Test and Kling-Eval. According to the model architecture, these works are categorized into diffusion- and autoregressive~(AR)-based methods, respectively. 

\textbf{VGGSound-Test.} VGGSound contains $\sim$15k video clips with 10s duration scratched from YouTube website. 
The original tags for these clips are brief phrases, which are inconsistent with our training annotation style. 
Thus, we adopt the designed data pipeline to generate audio captions for testing purposes.
As shown in Tab.~\ref{tab:sota_vggsound}, compared with the mainstream diffusion-based methods, the pioneer AR-based work, V-AURA~\cite{viertola2025temporally}, lags behind them by a large margin across multiple metrics. 
Through expanding the training data and designing a representative framework, DreamFoley has significantly improved the performance of AR-based models on this benchmark, meanwhile achieving results comparable to diffusion-based methods from various perspectives.
Notably, DreamFoley achieves competitive results in metrics related to fine-grained video-audio temporal alignment (\text{DeSync}), text-audio alignment (CLAP) and the generated data distribution (KL).

\textbf{Kling-Eval.} Kling-Eval includes $\sim$20k video clips with approximately 10s duration.
We synthesize the final audio description based on the video and audio caption provided in the official dataset.
We compare DreamFoley against contemporary methods in Tab.~\ref{tab:sota_kling}. 
ThinkSound~\cite{liu2025thinksound} leverages fine-grained textual supervision provided by LLM to guide audio generation. While effective at incorporating textual cues, its strong reliance on text introduces a significant bias. This compromises fidelity to the input video and leads to audio–visual misalignment, particularly in the absence of fine-grained audio descriptions.
MMAudio~\cite{cheng2025mmaudio} and HunyuanVideo-Foley~\cite{shan2025hunyuanvideo} both adopt a standard DiT-based diffusion architecture.
Notably, HunyuanVideo-Foley enhances the capability of diffusion model via expanding the training dataset to approximately 100,000 hours.
In comparison, DreamFoley achieves performance on par with HunyuanVideo-Foley across multiple evaluation metrics, and exhibits significant superiority in FD$_\text{PaSST}$, KL and CLAP scores.
These results demonstrate the potential of AR-based audio generation models, which surpasses diffusion-based approaches especially in instruction following tasks.

\definecolor{rowblue}{RGB}{235,242,247}
\definecolor{rowgray}{RGB}{192,192,192}

\begin{table}[t]
\centering
\footnotesize
\renewcommand{\arraystretch}{1.1}
\caption{Comparison with state-of-the-art methods on VGGSound-Test.}
\label{tab:sota_vggsound}
\setlength{\tabcolsep}{2pt}
\resizebox{\linewidth}{!}{
\begin{tabular}{
    l
    ccc
    c
    c
    c
    c
}
\toprule
\multirow{2}{*}{\textbf{Method}} & 
\multicolumn{3}{c}{\textbf{Distribution matching}} & 
\multicolumn{1}{c}{\textbf{Audio quality}} & 
\multicolumn{1}{c}{\textbf{Semantic align.}} & 
\multicolumn{1}{c}{\textbf{Temporal align.}} &
\multicolumn{1}{c}{\textbf{Text-Audio}} \\
\cmidrule(lr){2-4} \cmidrule(lr){5-5} \cmidrule(lr){6-6} \cmidrule(lr){7-7} \cmidrule{8-8}
 & FD$_{\text{PaSST}}\downarrow$ & FD$_{\text{PANNs}}\downarrow$ & KL$\downarrow$ & IS$\uparrow$ & IB-score$\uparrow$ & DeSync$\downarrow$  & CLAP$\uparrow$  \\
\midrule
\rowcolor{red!20}
\multicolumn{8}{c}{\textit{Diffusion-based methods}}\\
Frieren~\cite{wang2024frieren} & 106.10 & 11.45 & 2.79 & 12.25 & 22.78 & 0.85 & 0.11 \\
FoleyCrafter~\cite{zhang2024foleycrafter} & 140.09 & 16.24 & 2.26 & 15.68 & 25.68 & 1.23 & 0.19 \\
MMAudio~\cite{cheng2025mmaudio} & \textbf{60.60} & \textbf{4.72} & 1.53 & \textbf{17.40} & 33.22 & \textbf{0.44} & \textbf{0.25}\\
ThinkSound(\textit{w/o.} CoT)~\cite{liu2025thinksound} & 67.18 & 8.46 & \textbf{1.50} & 11.11 & 24.00 & 0.57 & 0.16\\
HunyuanVideo-Foley~\cite{shan2025hunyuanvideo} & 145.22 & 11.34 & 2.14  & 16.14 & \textbf{36.00} & 0.53 & 0.24\\ 
\rowcolor{red!20}
\multicolumn{8}{c}{\textit{Autoregressive-based methods}}\\
V-AURA~\cite{viertola2025temporally} & 218.50 & 14.80 & 2.25 & 10.08 & 27.64 & 0.65 & 0.12 \\
\rowcolor{rowblue}
DreamFoley (\textit{Ours})& \textbf{83.48} & \textbf{5.69} & \textbf{1.15} & \textbf{15.33} & \textbf{33.70} & \textbf{0.52} & \textbf{0.34} \\
\bottomrule
\end{tabular}
}
\end{table}

\begin{table}[t]
\centering
\footnotesize
\renewcommand{\arraystretch}{1.1}
\caption{Comparison with state-of-the-art methods on Kling-Eval.}
\label{tab:sota_kling}
\setlength{\tabcolsep}{2pt}
\resizebox{\linewidth}{!}{
\begin{tabular}{
    l
    ccc
    c
    c
    c
    c
}
\toprule
\multirow{2}{*}{\textbf{Method}} & 
\multicolumn{3}{c}{\textbf{Distribution matching}} & 
\multicolumn{1}{c}{\textbf{Audio quality}} & 
\multicolumn{1}{c}{\textbf{Semantic align.}} & 
\multicolumn{1}{c}{\textbf{Temporal align.}} &
\multicolumn{1}{c}{\textbf{Text-Audio}} \\
\cmidrule(lr){2-4} \cmidrule(lr){5-5} \cmidrule(lr){6-6} \cmidrule(lr){7-7} \cmidrule{8-8}
 & FD$_{\text{PaSST}}\downarrow$ & FD$_{\text{PANNs}}\downarrow$ & KL$\downarrow$ & IS$\uparrow$ & IB-score$\uparrow$ & DeSync$\downarrow$  & CLAP$\uparrow$  \\
\midrule
\rowcolor{red!20}
\multicolumn{8}{c}{\textit{Diffusion-based methods}}\\
Frieren~\cite{wang2024frieren} & 293.57 & 16.86 & 2.95 & 7.32 & 21.0 & 0.86 & 0.16 \\
FoleyCrafter~\cite{zhang2024foleycrafter} & 322.63& 22.30 & 2.47 & 7.08 & 22.0 & 1.23 & 0.22 \\
MMAudio~\cite{cheng2025mmaudio} & 205.85 & 9.01 & 2.17 & \textbf{9.59} & 30.0 & 0.56 & \textbf{0.27} \\
ThinkSound(\textit{w/o.} CoT)~\cite{liu2025thinksound} & 228.68 & 9.92 & 2.39 & 6.86 & 22.0 & 0.67 & 0.22 \\
HunyuanVideo-Foley~\cite{shan2025hunyuanvideo} & \textbf{202.12} & \textbf{6.07} & \textbf{1.89} & 8.30 & \textbf{38.0} & \textbf{0.54} & 0.24\\ 
\rowcolor{red!20}
\multicolumn{8}{c}{\textit{Autoregressive-based methods}}\\
V-AURA~\cite{viertola2025temporally} & 474.56 & 33.15 & 3.24 & 5.80 & 25.0 & 0.86 & 0.13 \\
\rowcolor{rowblue}
DreamFoley (\textit{Ours}) & \textbf{127.73} & \textbf{6.88} & \textbf{1.65} & \textbf{8.63} & \textbf{31.2} & \textbf{0.60} & \textbf{0.32} \\
\bottomrule
\end{tabular}
}
\end{table}

\subsection{Ablation Study}
In this section, we perform several ablation studies to evaluate the importance of key components in our framework. 
For fair comparison, all experiments are conducted on the VGGSound-Test dataset across various metrics.

\textbf{Effect of dual vision encoder.} In Tab.~\ref{tab:vision_enc}, we compare the performance of different visual encoders.
This experiment is conducted without the CFG strategy to isolate the contribution of the encoders.
Specifically, as indicated by the DeSync metric, the audio-aligned branch delivers more precise information for audio–video synchronization.
In contrast, the text-aligned branch focuses on extracting semantic information from visual contents.
Notably, the combination of both branches achieves the best performance, simultaneously improving the quality of generated audio and ensuring optimal synchronization between audio and video.

\textbf{Effect of RVQ layer number.} In Tab.~\ref{tab:rvq_layer}, we evaluate how the number of RVQ layers affects model performance.
These results reveal that using only the first layer yields the weakest performance across all metrics, indicating limited quantization capacity at shallow levels. 
Expanding to four layers improves results significantly, demonstrating the benefits of incorporating deeper quantization. 
Notably, using all eight layers achieves the best performance, with the lowest KL, highest IS, and peak IB-score. Metrics like DeSync and CLAP also show consistent improvement.

\textbf{Effect of CFG scale.} In Tab.~\ref{tab:cfg_scale}, we investigate the effect of CFG on model performance by varying the guidance scale. The results demonstrate that without CFG ($\gamma=1$), the model performs the worst across all metrics. Introducing CFG significantly improves the results, while the performance peaks at $\gamma=3$, reflecting a balance between diversity and alignment in the generated outputs. However, further increasing the scale slightly degrades performance, likely due to over-amplifying the guidance signal, which could distort the sampling distribution. 

\begin{table}[htbp]
  \centering
  \begin{minipage}{0.45\textwidth}
    \centering
    \footnotesize
    \renewcommand{\arraystretch}{1.1}
    \caption{Effect of dual vision encoder.}
    \label{tab:vision_enc}
    \setlength{\tabcolsep}{4pt}
    \resizebox{\linewidth}{!}{
    \begin{tabular}{
        l
        c
        c
        c
        c
        c
    }
    \toprule
    \multirow{2}{*}{\textbf{Vision Encoder}} & 
    \multicolumn{5}{c}{\textbf{VGGSound-Test}} \\
    \cmidrule(lr){2-6}
     & KL$\downarrow$ & IS$\uparrow$ & IB-score$\uparrow$ & DeSync$\downarrow$  & CLAP$\uparrow$  \\
     \midrule
     Audio-aligned & 1.53 & 10.64 & 25.55 & 0.72 & 0.29 \\
     Text-aligned & 1.57 & 10.43 & 25.59 & 1.09 & \textbf{0.30} \\
     Audio+Text-aligned & \textbf{1.49} & \textbf{10.88} & \textbf{26.54} & \textbf{0.71} & 0.29\\
     \bottomrule
    \end{tabular}
    }
  \end{minipage}
  \hspace{0.03\textwidth} 
  \begin{minipage}{0.45\textwidth}
    \centering
    \footnotesize
    \renewcommand{\arraystretch}{1.1}
    \caption{Effect of RVQ layer number.}
    \label{tab:rvq_layer}
    \setlength{\tabcolsep}{4pt}
    \resizebox{\linewidth}{!}{
    \begin{tabular}{
        l
        c
        c
        c
        c
        c
    }
    \toprule
    \multirow{2}{*}{\textbf{RVQ layer}} & 
    \multicolumn{5}{c}{\textbf{VGGSound-Test}} \\
    \cmidrule(lr){2-6}
     & KL$\downarrow$ & IS$\uparrow$ & IB-score$\uparrow$ & DeSync$\downarrow$  & CLAP$\uparrow$  \\
     \midrule
     Layer 1 & 1.71 & 7.09 & 19.28 & 0.74 & 0.27 \\
     Layer 1-4 & 1.56 & 9.42 & 24.39 & 0.71 & 0.29 \\
     Layer 1-8 & \textbf{1.49} & \textbf{10.88} & \textbf{26.54} & \textbf{0.71} & \textbf{0.29} \\
     \bottomrule
    \end{tabular}
    }
  \end{minipage}
\end{table}

\begin{table}[htbp]
  \centering
  \begin{minipage}{0.45\textwidth}
    \centering
    \footnotesize
    \renewcommand{\arraystretch}{1.1}
    \caption{Effect of CFG scale $\gamma$.}
    \label{tab:cfg_scale}
    \setlength{\tabcolsep}{4pt}
    \resizebox{\linewidth}{!}{
    \begin{tabular}{
        c
        c
        c
        c
        c
        c
    }
    \toprule
    \multirow{2}{*}{\textbf{CFG scale}} & 
    \multicolumn{5}{c}{\textbf{VGGSound-Test}} \\
    \cmidrule(lr){2-6}
     & KL$\downarrow$ & IS$\uparrow$ & IB-score$\uparrow$ & DeSync$\downarrow$  & CLAP$\uparrow$  \\
     \midrule
     1 & 1.58 & 10.75 & 26.1 & 0.72 & 0.27 \\
     2 & 1.34 & 13.27 & 30.2 & 0.66 & 0.31 \\
     3 & \textbf{1.15} & \textbf{15.33} & \textbf{33.7} & \textbf{0.52} & \textbf{0.34} \\
     4 & 1.24 & 14.13 & 32.6 & 0.57 & 0.32 \\
     \bottomrule
    \end{tabular}
    }
  \end{minipage}
  \hspace{0.03\textwidth} 
  \begin{minipage}{0.45\textwidth}
    \centering
    \footnotesize
    \renewcommand{\arraystretch}{1.1}
    \caption{Effect of various audio tokenizers.}
    \label{tab:tokenizer}
    \setlength{\tabcolsep}{4pt}
    \resizebox{\linewidth}{!}{
    \begin{tabular}{
        l
        c
        c
        c
        c
        c
    }
    \toprule
    \multirow{2}{*}{\textbf{Tokenizer}} & 
    \multirow{2}{*}{\textbf{Type}} &
    \multicolumn{4}{c}{\textbf{VGGSound-Test}} \\
    \cmidrule(lr){3-6}
     & & KL$\downarrow$ & IS$\uparrow$ & IB-score$\uparrow$ & DeSync$\downarrow$ \\
     \midrule
     \rowcolor{rowgray}
     GT & -- & 0 & 14.8 & 33.22 & 0.63  \\
     \cmidrule(lr){1-6}
     Xcodec~\cite{ye2025codec} & Discrete & 0.47 & 9.3 & 26.91 & 0.71  \\
     Higgs-Audio & Discrete & 0.31 & 11.1 & 28.69 & 0.68 \\
     \cmidrule(lr){1-6}
     MMAudio~\cite{cheng2025mmaudio} & VAE & 0.14 & 13.2 & 31.49 & 0.65 \\
     \bottomrule
    \end{tabular}
    }
  \end{minipage}
\end{table}

\subsection{Qualitative Results}
\begin{figure}[t]
	\centering
	\includegraphics[width=1.0\linewidth]{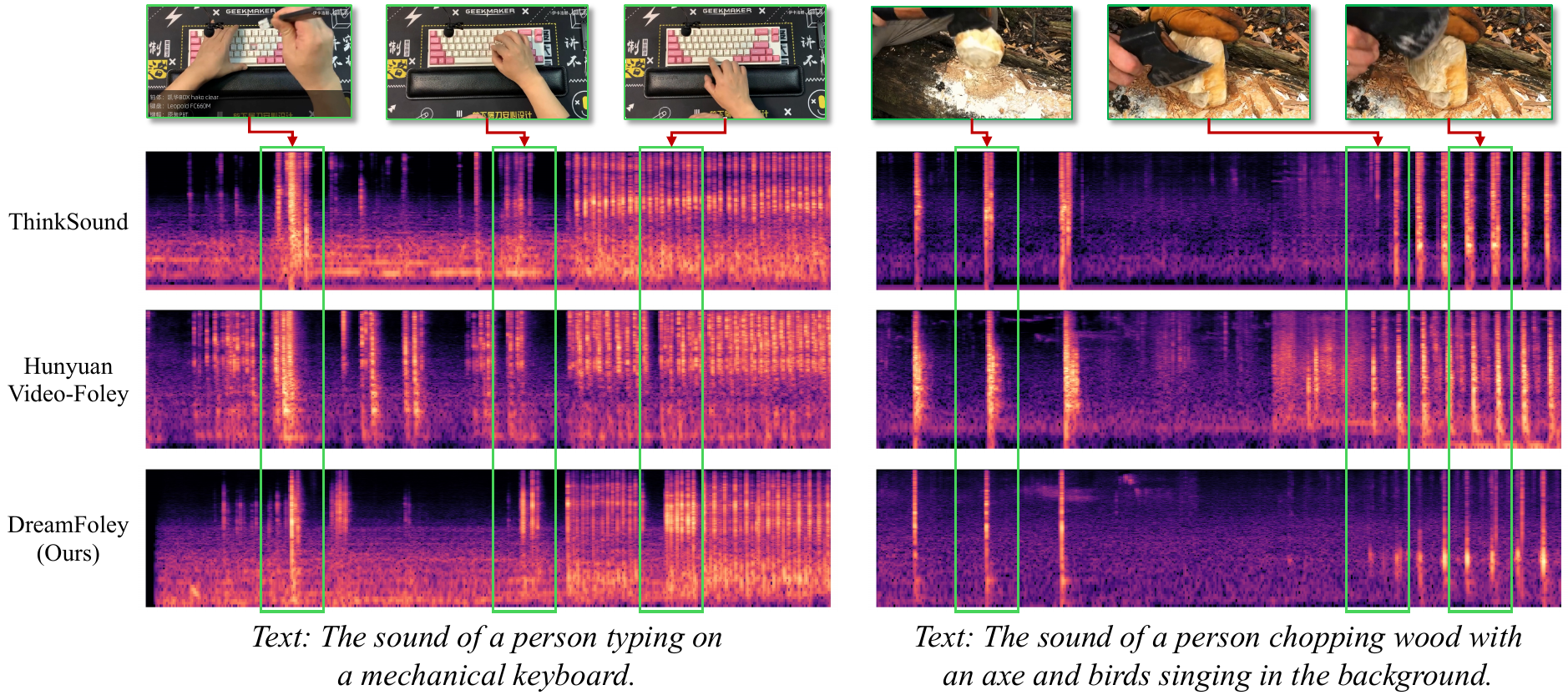} 
	\caption{Qualitative comparison between DreamFoley and previous methods, \textit{i.e.,} ThinkSound and HunyuanVideo-Foley. The samples are derived from the Kling-eval dataset. The audio descriptions are generated from data production pipeline.} 
	\label{fig:visualize}
\end{figure}

We visualize the spectrograms of the audio generated by DreamFoley and prior works, as shown in Fig.~\ref{fig:visualize}. 
It is evident that the audio generated by DreamFoley aligns more accurately with the sound-producing events depicted in the video, while also exhibiting lower levels of background noise. For example, in the sample shown on the right, the wood-chopping sound produced by DreamFoley increases gradually in intensity, corresponding more appropriately to the degree of exertion displayed by the person in the video. In contrast, audio generated by other methods appears overly rigid and lacks realism.

\section{Conclusion}

In this paper, we introduced DreamFoley, a novel autoregressive framework for high-fidelity video-to-audio generation. 
By leveraging the powerful sequential modeling capabilities of large vision-language models (VLMs), our approach effectively addresses key limitations of prior diffusion-based methods, particularly in terms of temporal scalability and semantic alignment.
The core innovations of DreamFoley include a dual-visual encoder architecture that captures both fine-grained temporal cues and global semantic information, a Residual Vector Quantization (RVQ) audio tokenizer with a delay-pattern generation scheme for efficient and high-quality audio representation, and the strategic application of CFG within an autoregressive framework to enhance output fidelity and alignment.
Through extensive experiments on popular benchmarks such as VGGSound-Test and Kling-Eval, DreamFoley demonstrates competitive or superior performance compared to state-of-the-art diffusion-based and autoregressive methods.
Looking forward, we believe that AR-based models like DreamFoley offer a promising direction for scalable and high-quality video-to-audio synthesis. 

Despite DreamFoley demonstrating that AR-based models perform on par with diffusion-based approaches for audio generation tasks, our study identifies several factors that may limit the upper bound of AR model performance, among which the audio tokenizer is particularly critical. 
In diffusion-based methods, the prevailing practice is to employ VAE-based continuous feature compression for audio tokenization. By contrast, discrete audio tokenizers inevitably suffer from quantization errors. Even with techniques like RVQ, they still exhibit a substantial gap in reconstruction quality compared to VAE-based approaches.
To validate this hypothesis, we reconstructed the ground-truth audio from VGGSound-Test and compared the metrics across different tokenization schemes. As shown in Tab.~\ref{tab:tokenizer}, the reconstructed audio produced by discrete tokenizers underperforms VAE-based reconstructions across all evaluated metrics. This finding indicates that improving the reconstruction fidelity of the audio tokenizer has a direct and significant impact on the attainable ceiling of audio generation quality.
In addition, the current implementation of DreamFoley supports audio generation for videos up to 10 seconds in length. Nevertheless, due to the inherently causal architecture of AR models, the approach can be extended seamlessly to substantially longer audio durations.
Overall, future work will focus on extending the framework to support longer audio generation, enhancing the audio tokenizer, and exploring more advanced conditioning mechanisms for even finer-grained audio-visual synchronization.

\clearpage

\bibliographystyle{plainnat}
\bibliography{moment_main}




\end{document}